\documentclass[10pt, conference, compsocconf]{IEEEtran}
  
\usepackage{graphicx}
\usepackage{url}
\usepackage{amsmath,amsfonts}
\usepackage{amsthm} 
\usepackage{mathtools}
\usepackage{color}
\usepackage{xspace}
\usepackage{cite}

\usepackage{enumitem}

\setlist[enumerate]{noitemsep,topsep=0pt,parsep=1pt,partopsep=1pt}

\theoremstyle{definition}
\newtheorem{theorem}{Theorem}
\newtheorem{lemma}{Lemma}

\newtheorem{definition}{Definition}
\newtheorem{innercustomrule}{Rule}
\newenvironment{customrule}[1]
	{\renewcommand\theinnercustomrule{#1}\innercustomrule}
	{\endinnercustomrule}
    
\renewenvironment{quote}
  {\list{}{\leftmargin=0.1in\rightmargin=0.1in}\item[]}%
  {\endlist}

\newcommand{\cutoff}[1]{}  
\newcommand{\shortonly}[1]{}
\newcommand{\extendedonly}[1]{#1}
\newcommand{\AG}[4]{\left< #3 \right> #1 \left[ #2 \right] \left< #4 \right>}
\newcommand{\myparagraph}[1]{\vspace*{0.06in}\noindent\textbf{#1}}

\newcommand{\articleonly}[1]{}

\newcommand{\scriptC}[1][]{\ensuremath{\mathcal{C}_{#1}}\xspace}
\newcommand{\QB}{{\it QB}\xspace}		
\newcommand{\MP}{{\it MP}\xspace}		
\newcommand{\Nav}{{\it Nav}\xspace}	
\newcommand{\Plant}{{\it ILP}\xspace}	
\newcommand{\PS}{{\it PS}\xspace}		
\newcommand{\ES}{{\it ES}\xspace}		
\newcommand{\CF}{{\it CF}\xspace}		
\newcommand{\BE}{{\it BE}\xspace}		
\newcommand{\FE}{{\it FE}\xspace}		
\newcommand{\BT}{{\it BT}\xspace}		
\newcommand{\MC}{{\it MC}\xspace}		
\newcommand{\TU}{{\it TU}\xspace}		

\newcommand{\note}[1]{}

\newcommand{\intersect}{\cap}
\newcommand{\union}{\cup}

\DeclarePairedDelimiter\abs{\lvert}{\rvert}%
\newcommand{\degree}{$^\circ$\xspace}
\renewcommand{\implies}{\;\Rightarrow\;}

\begin{document}

\title{A Component-Based Simplex Architecture for\\ High-Assurance Cyber-Physical Systems}

\author{
	\IEEEauthorblockN{Dung~Phan\IEEEauthorrefmark{1}, Junxing~Yang\IEEEauthorrefmark{1}, 
    	Matthew~Clark\IEEEauthorrefmark{2}, Radu~Grosu\IEEEauthorrefmark{3}, 
        John~Schierman\IEEEauthorrefmark{4}, Scott~Smolka\IEEEauthorrefmark{1}
        and Scott~Stoller\IEEEauthorrefmark{1}
    }
	\IEEEauthorblockA{
    	\IEEEauthorrefmark{1}Department of Computer Science\\
			Stony Brook University, Stony Brook, NY, USA}
		\IEEEauthorblockA{\IEEEauthorrefmark{2}Air Force Research Laboratory, Dayton, OH, USA}
		\IEEEauthorblockA{\IEEEauthorrefmark{3}Department of Computer Science\\
        	Vienna University of Technology, Vienna, Austria}
		\IEEEauthorblockA{\IEEEauthorrefmark{4}Barron Associates Inc., Charlottesville, VA, USA}
}

\maketitle
\begin{abstract}

We present \emph{Component-Based Simplex Architecture} (CBSA), a new
framework for assuring the runtime safety of
component-based cyber-physical systems (CPSs).  CBSA integrates Assume-Guarantee
(A-G) reasoning with the core principles of the Simplex control architecture to allow
component-based CPSs to run advanced, uncertified controllers while still providing runtime assurance that A-G contracts and global properties are satisfied.  
In CBSA, multiple Simplex instances, which can be composed in a nested, 
serial or parallel manner, coordinate to assure system-wide properties.

Combining A-G reasoning and the Simplex architecture is a challenging problem that yields significant benefits.
By utilizing A-G contracts, we are able to \emph{compositionally} determine the switching logic for CBSAs, thereby alleviating 
the state explosion encountered by other approaches.
Another benefit is that 
we can use A-G proof rules to decompose 
the proof of system-wide safety assurance into sub-proofs corresponding to
the component-based structure of the system architecture.
We also introduce the notion of 
\emph{coordinated switching} between Simplex instances, a key component of our compositional approach to reasoning about CBSA switching logic.

We illustrate our framework with a component-based control
system for a ground rover.
We formally prove that the CBSA for this system 
guarantees \emph{energy safety} (the rover never runs out of power), and \emph{collision freedom} (the rover never collides with a stationary obstacle).  We also consider a CBSA for the rover that guarantees \emph{mission completion}: all target destinations
visited within a prescribed amount of time.  

\end{abstract}
\section{Introduction} 
\label{sec:intro}

\begin{figure}[htbp] 
	\begin{center} 
		\includegraphics[scale=0.5]{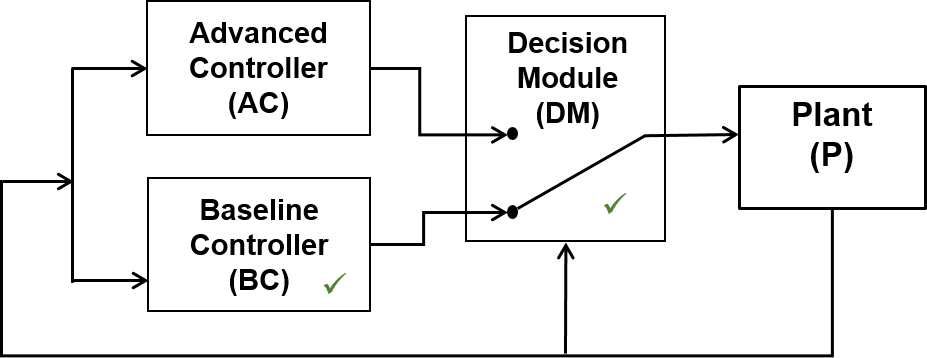} 
		\caption{The Simplex architecture.  The Decision Module 
			and Baseline Controller are pre-certified.} 
		\label{fig:simplex} 
	\end{center} 
    \vspace*{-0.1in}
\end{figure}

Simplex~\cite{seto98simplex,seto98dynamic,sha01using} is a software 
architecture for high-assurance process-control systems.  It traditionally 
consists of a physical plant and two versions of the controller:
an \emph{advanced controller} (AC) and a \emph{baseline controller}
(BC).  The AC is in control of the plant under nominal operating
conditions, and is designed to achieve \emph{high-performance} according
to certain metrics (e.g., maneuverability, fuel economy,
mission-completion time).  The BC is pre-certified to keep the plant
within a prescribed \emph{safety region}, i.e., a region of safe 
operation.  A \emph{decision module} (DM), which is also pre-certified,
continually monitors the state of the plant and switches control of the
plant to the BC should the plant be in imminent danger (i.e., within the
next update period) of exiting the safety region.  As such, Simplex is 
a very powerful architecture.  It assures that the plant is properly 
controlled even if the advanced controller has bugs.  
As advanced controllers are increasingly more complex, more adaptive
with the use of unverified algorithms such as machine-learning's, 
runtime assurance techniques like Simplex are becoming more important.
Figure~\ref{fig:simplex} illustrates the Simplex architecture.

The Simplex architecture was developed 20 years ago.  
Since then,
control systems
have evolved to take on a more complex
structure consisting of multiple controllers, each with a distinct
functionality.  For example, an autonomous vehicle might have a
hierarchical control system consisting of  
controllers for (in order of decreasing level of abstraction) 
mission-planning, guidance, navigation, and inner-loop control.  A system
of this form is illustrated in Fig.~\ref{fig:quickbot-simplex}.  
Furthermore, some of these controllers themselves may have a hierarchical 
or nested structure.

Advanced control systems are not necessarily hierarchical, 
but are better understood as having a \emph{component-based} 
architecture.   A ``component'' may be a software module
or the distinguished physical-plant node (we do not consider multiple 
physical plants, because a physical plant can be arbitrarily complex).
It is important to have runtime assurance techniques for
component-based control systems that are equally modular, to
reduce their associated complexity and cost.

Components may have different \emph{update periods} (or \emph{execution
periods}).  For the autonomous-vehicle example, 
the mission-planning component's update period may be
a multiple of navigation's, which in turn may be a multiple of inner-loop's.  
The composition of these components is a \emph{multi-rate} component. 
Our framework allows this multi-rate characteristic to be modeled explicitly.

Adapting the Simplex architecture to component-based systems is non-trivial.
On the one hand, wrapping a large monolithic Simplex instance around the entire 
control system violates modularity principles, making the design and verification 
of the system more difficult.  On the other hand, naively wrapping each
component of the system in an independent Simplex container is inadequate to ensure 
system-wide properties, because of interactions between components.

For example, to guarantee that a ground rover never runs out of power,
the mission-planning controller might change the current target to the
most recently observed power station when it detects that the battery 
level is low.  It needs to rely on specific properties of other components to
ensure that the rover can reach the power station in question before
depleting the battery, and these properties might not be assured by the advanced 
controllers of those components.  
This example also highlights the need for a compositional proof technique, 
since a property is assured not by a single Simplex instance, but rather by multiple
cooperating instances.

Assume-Guarantee (A-G) reasoning~\cite{abadi95conjoining,mcmillan97compositional,stark85rely}
is a powerful compositional proof technique.  With this approach,
to prove that a system composed of components $M_1$ and $M_2$
satisfies a property $q$, one shows that $M_1$ satisfies $q$ under
assumptions $p$ and that $M_2$ satisfies $p$.  Assumption $p$ 
and guarantee $q$ of $M_1$ are specified in $M_1$'s \emph{A-G contract}.
A-G contracts specify a component's behavior for a single time step, and inductive
reasoning about contracts is used to prove system-wide invariants.
Furthermore, this approach is applicable to arbitrary safety properties, 
as a safety property can be expressed as an invariant by adding auxiliary 
variables~\cite{schneider97concurrent}.

\myparagraph{Contributions.}
In this paper, we present \emph{Component-Based Simplex Architecture} 
(CBSA), a new framework for assuring the runtime safety of
component-based cyber-physical systems.
CBSA integrates Assume-Guarantee reasoning with the core principles of
the Simplex control architecture to allow component-based control systems 
to run advanced, uncertified controllers while still being able
to provide runtime assurance that A-G contracts, and the global invariants 
they imply, are satisfied.  

A novelty of CBSA is that by utilizing A-G contracts, 
we are able to \emph{compositionally} determine 
Simplex switching logic, thereby alleviating the state explosion problem 
encountered by other approaches.  
Prior methods~\cite{bak2014rtss,bak11sandboxing,seto99acase} 
for deriving Simplex switching logic 
are not designed to be modular,
so components 
would need to be composed into a monolithic system before applying these approaches. 
In contrast, we compositionally derive Simplex switching logic 
for a component using assumptions it makes about its environment.
These assumptions are then discharged by the guarantees specified 
in A-G contracts of other components.

Another feature of CBSA is its use of A-G reasoning as a 
proof technique for runtime assurance.  In CBSA, Simplex 
instances are used to guarantee local, component-based 
A-G contracts.  These contracts specify, for each component, 
the assumptions it makes about its environment and 
the guarantees it provides if the environment satisfies 
those assumptions.  A-G reasoning, based on 
A-G proof rules, is used at compile time to provide 
runtime assurance of global safety properties from the 
locally assured A-G contracts.

CBSA also introduces the notion of \emph{coordinated switching} among 
Simplex-based components.  This is needed when the violation of a 
contract in one component leads to a cascade of contract violations 
in other components, thereby necessitating  synchronized switching to 
ensure that the desired system properties are not violated.


Another contribution of the paper is a detailed case study that 
thoroughly illustrates the principles of CBSA.  In particular, 
we show how to apply our framework to a component-based control
system for the \emph{Quickbot ground rover}~\cite{quickbot14}.
The Quickbot control system includes a \emph{Mission Planning}
component, which selects targets (i.e., destinations), and a \emph{Navigation} 
component, which steers the rover to its next target destination.

We use A-G reasoning to formally prove that the CBSA for this system
guarantees two safety properties: (i)~\emph{energy safety} (ES), 
which ensures the rover never runs out of power, and
(ii)~\emph{collision freedom} (CF), which ensures the rover never collides
with an obstacle.  This example illustrates the unique
characteristics of CBSA.  The switching logic in all Simplex instances
is designed to guarantee component-based A-G contracts.  
Moreover, coordinated switching between the Mission Planning and
Navigation components is required to ensure ES.

The ES property illustrates how we use A-G contracts
to compositionally construct Simplex switching logic.  
To ensure energy safety, the rover backtracks to the most recently
visited power station when the battery level drops below a threshold.
We compositionally derive this threshold for the \emph{Mission Planning}'s 
switching logic by using a guarantee in the contract of 
the \emph{Navigation} component.  This guarantee assures that the
energy the rover needs to backtrack to the most
recently visited power station will not exceed the forward energy 
it has expended to travel from this power station to
the current position.  

\emph{Navigation} assures this guarantee
by using a Simplex instance that must switch in tandem with
\emph{Mission Planning}'s instance.  Therefore, we enforce a 
\emph{coordinated switching} between \emph{Mission Planning} 
and \emph{Navigation} to ensure the satisfaction of both components'
A-G contracts and the global \emph{energy safety} property they
imply.

We additionally consider a \emph{Mission Completion} (MC) property 
for the Quickbot system, i.e., all target destinations are  visited 
within a prescribed amount of time.  This property is interesting 
not only because it is a (bounded) liveness property, a type of property 
not traditionally handled by Simplex, but also because it is a 
higher-level mission-oriented property several layers of control removed 
from the physical plant.  The MC property is also more software-oriented
in nature, as it revolves around a software data structure (a list of mission targets).

The rest of the paper is organized as follows.  Section~\ref{sec:related} considers
related work.  Section~\ref{sec:cbsa} formally introduces our CBSA framework.
Sections~\ref{sec:case-study} and~\ref{sec:MC} present the Quickbot case studies. 
Section~\ref{sec:conclusion} offers 
our concluding remarks and directions for future work.

\section{Related Work}
\label{sec:related}

The relationship of our work with the classical Simplex architecture is 
discussed in Section~\ref{sec:intro}.

Schierman {\it et al}.\ developed a runtime assurance technique similar to Simplex, known simply as RTA~\cite{aiello10runtime,schierman08runtime}.  
RTA can be applied to component-based systems, but each RTA wrapper (i.e., each 
Simplex-like instance) independently ensures a local safety property of a component.  
For example, in~\cite{aiello10runtime}, RTA instances for an inner-loop controller 
and a guidance system are uncoordinated and thereby operate independently. Their work does not 
consider A-G contracts or cooperation between components to ensure global properties
like we do in this paper.


Schierman {\it et al}.\ recently extended their work on RTA by showing how it can be used to ensure that components satisfy A-G contracts~\cite{schierman15}. They apply RTA to, and give A-G contracts for, 
several components in an Unmanned Aerial Vehicle (UAV) flight control system.  They do not use formal 
A-G proof rules to derive global invariants.  Instead, they reason about the 
overall safety of the system using Goal-Structuring Notation (GSN), which is 
less formal.  They discuss the potential need for coordinated switching, 
but coordinated switching is not used in their case study.

A-G reasoning has been extensively studied for the
compositional verification of complex systems 
(e.g., \cite{abadi95conjoining,alur99reactive,chandy88parallel,clarke89compositional,grumberg94model,mcmillan97compositional,pnueli89in,stark85rely}).  
There is also significant tool support for A-G reasoning, including
AGREE~\cite{cofer12compositional}, OCRA~\cite{cimatti13ocra}, 
and Safety ADD~\cite{warg14safety}. 
AGREE is a framework that supports A-G reasoning on 
architectural models specified in AADL.
OCRA supports the specification of A-G contracts using 
temporal logic and can also be used to verify the correctness of contract refinements.
Safety ADD is 
a tool for defining A-G contracts and verifying that all guarantee-assumption pairs match and there are no unfulfilled assumptions.
A-G reasoning has also been used for
compositionally checking source code for preservation of the design's
correctness~\cite{giannakopoulou04assume}.

In~\cite{wang2016multi}, compositional barrier functions are used to provably guarantee the simultaneous satisfaction of composed objectives. They rely on a single controller and an optimization based approach to correct the controller when violations of safety are imminent, which has limited capability and less flexibility compared to Simplex.

None of these approaches, however, consider the possibility of pairing A-G reasoning
with techniques for runtime assurance of system properties.
Our approach integrates A-G reasoning with the core principles of Simplex, 
allowing component-based
systems to run advanced, uncertified controllers while still being able
to provide runtime assurance that A-G contracts, and the global invariants 
they imply, are satisfied.

A comparison between our work and the Simplex switching-logic-derivation techniques in~\cite{bak2014rtss,bak11sandboxing,seto99acase} 
is given in Section~\ref{sec:intro}.
Without A-G contracts and our concept of coordinated switching, 
those approaches could be extended as follows to apply to systems with multiple
Simplex instances.  Suppose there are $n$ Simplex instances, and $s$ is the decision period.  
A global decision module can choose among $2^n$ possible controller configurations 
(baseline or advanced for each Simplex instance).  
For each of these $2^n$ configurations, do backward reachability 
analysis~\cite{bak11sandboxing} with an unbounded time horizon for the system 
composed from the selected controllers and the plant, 
using maximally nondeterministic models for the advanced controllers,
to obtain a recoverable region for that configuration.
At each decision time, the global decision module checks whether 
(1)~all states reachable in time at most $s$ using the current controller configuration are safe, and 
(2)~all states reachable in time exactly $s$ using the current controller configuration are in the recoverable region of the current controller configuration.  If so, the decision module continues to use the current configuration; otherwise, it switches to a controller configuration for which these two conditions hold.

\section{Component-Based Simplex Architecture}
\label{sec:cbsa}


\subsection{Multi-rate Components}
\label{sec:components}
\begin{definition}
\label{def:component}
{ 
    \def\OldComma{,}
    \catcode`\,=13
    \def,{%
      \ifmmode%
        \OldComma\discretionary{}{}{}%
      \else%
        \OldComma%
      \fi%
    }%
A \emph{multi-rate component} $M$ is a tuple $(x, u, y, S)$, 
where 
$x = \{x_1, ..., x_m\}$ is the set of \emph{state variables},
$u = \{u_1, ..., u_k\}$ is the set of \emph{input variables},
$y = \{y_1, ..., y_n\}$ is the set of \emph{output variables},
and $S = \{(f_1, g_1, s_1), ..., (f_l, g_l, s_l)\}$
is a sequence whose $i$-th element is a triple of a \emph{next-state function} $f_i$, 
an \emph{output function} $g_i$, and their \emph{update period} $s_i$, which is a positive
integer multiplier of the global clock tick $dt$.  The sets of input and output variables
are disjoint.  The behavior of $M$ at tick $i$
is defined as follows.
}
\begin{equation*}
\left\{
\begin{array}{lcl}
  x(i) = f_j\left(x(i-1), u(i)\right),  \forall j\in [1..l]: 	i \text{ mod } s_j = 0\\
  y(i) = g_j\left(x(i), u(i)\right),  \forall j\in [1..l]: 	i \text{ mod } s_j = 0\\
\end{array}
\right.
\end{equation*}
\noindent
where the set assignment $A = B$ means only the variables in $A \intersect B$  are assigned the corresponding values in $B$.  The variables in $A$ that
are not updated at tick $i$ retain the values they had at tick $i-1$.  
\end{definition}

We give examples of components in Section~\ref{sec:quickbot-cbsa}.


\subsection{Composition with Feedback}
\label{sec:serial-composition}

For simplicity, we make the following assumptions about multi-rate components.
\begin{enumerate}
	\item Components communicate via shared variables.  After a 
	component writes a shared variable, the updated value is instantly
	available to be read by all components.
	\item If a component writes and another component reads a shared 
	variable at the same tick, then there must be a 
	predetermined order of execution of the two components. 
    The order of execution of $M_1 \parallel M_2$ is $M_1$, $M_2$.
    This means
    the composition operator $\parallel$ is associative but not necessarily commutative.
	\item The execution time of a component is negligible.
    \item { 
    \def\OldComma{,}
    \catcode`\,=13
    \def,{%
      \ifmmode%
        \OldComma\discretionary{}{}{}%
      \else%
        \OldComma%
      \fi%
    }%
    A 1-tick delay is introduced in feedback loops to break circularity.  A feedback loop exists between two components 
    $M_1 = (x_1, u_1, y_1, S_1)$ and $M_2 = (x_2, u_2, y_2, S_2)$ if 
    $y_1 \intersect u_2 \neq \emptyset$ and $y_2 \intersect u_1 \neq \emptyset$.
    If the order of execution is $M_1$, $M_2$ then the variables in $y_2$ that feed
    back into $u_1$ are delayed by one tick; i.e., $y_2(i-1)$
    is supplied to $f_i$ and $g_i$ of $M_1$ instead of $y_2(i)$.  }
\end{enumerate}

{ 
    \def\OldComma{,}
    \catcode`\,=13
    \def,{%
      \ifmmode%
        \OldComma\discretionary{}{}{}%
      \else%
        \OldComma%
      \fi%
    }%
Two components $M_1 = (x_1, u_1, y_1, S_1)$ and 
$M_2 = (x_2, u_2, y_2, S_2)$
are \emph{composable} if $x_1 \intersect x_2 = \emptyset$ and $y_1 \intersect y_2 = \emptyset$.
\begin{definition}
\label{def:composition}
Let $M_1$ and $M_2$ be composable.  Then their \emph{composition} $M_1 \parallel M_2$ is the component
$(x, u, y, S)$, 
where $x = x_1 \union x_2 \union (y_1 \intersect u_2) \union (y_2 \intersect u_1)$, 
$u = (u_1 \union u_2) - (y_1 \union y_2)$,
$y = y_1 \union y_2$, and $S = S_1 \union S_2$.
\end{definition}
Functions $f_i$ and $g_i$ of $M_1 \parallel M_2$ ignore variables in $x$ and $u$
that are not their original arguments.
Note that the connected inputs
$(y_1 \intersect u_2) \union (y_2 \intersect u_1)$ are hidden from $u$ and added to $x$.
This is necessary to ensure functions
$f_i, g_i$ can still access their required variables.
The variables in $x  \intersect  y$ are synchronized and updated by functions $g_i$.
Because the order of execution is $M_1$, $M_2$, functions $f_i$ and $g_i$ in $S_1$ 
are executed before those in $S_2$.
}
\note{I think we already said if a variable is not updated at time tick i, 
it retains the value it had at time tick i-1. So the last sentence is 
confusing. I want to change it to "This is necessary to ensure functions
$f_i, g_i$ can still access their required variables." -dung}

\vspace{6pt}
\noindent
\textbf{Remark.} Our definition of component is similar to~\cite{lee2003structure}
except we have multi-rate.  Our form of non-commutative composition can be
regarded as a shorthand for a combination of cascade composition and feedback
composition in~\cite{lee2003structure}. 
For example, suppose component $M_1$ reads variable $x$ and writes variable $y$, 
and component $M_2$ reads $y$ and writes $x$.  To obtain a system equivalent to our composition 
$M_1 \parallel M_2$ using the combinators in~\cite{lee2003structure}, one can
first construct a cascade composition of $M_1$, $M_2$, and a unit delay state machine 
$M_D$ that delays $x$ by one tick, and then perform a feedback composition on
the resulting component.





\subsection{Assume-Guarantee Contracts}
\label{sec:ag-contracts}

To prove system properties in CBSA, we use A-G reasoning in the form of
A-G contracts~\cite{benveniste2012contracts}.  That is, with every component $M$ of a 
CBSA, we associate a contract of the form $\scriptC = (I,O,A,G)$, where $I$ is a 
set of typed \emph{input} variables, $O$ is a set of typed 
\emph{output} variables, $A$ is an \emph{assumptions} predicate, and 
$G$ is a \emph{guarantees} predicate.  Note that a contract cannot
guarantee a system-wide property 
involving a (global) variable that does not belong to the contract.  A 
contract should only talk about the assumptions and guarantees of its 
inputs and outputs.  A-G contracts allow \emph{compositional reasoning}
about CBSAs using the A-G reasoning rule given in Section~\ref{sec:ag-rule}.  
In CBSA, the switching logic for the Simplex instance of a component 
is based on determining whether the active AC might violate the
component's guarantees in the next decision period.  
Furthermore, the composition of all active controllers must
imply the safety properties of the system.  This principle results in
a requirement that the composition of all BCs in a CBSA must imply
the safety properties of the system in case all components must switch
to their BCs.

\subsection{Assume-Guarantee Proof Rule}
\label{sec:ag-rule}

An \emph{assume-guarantee triple} of the form $\AG{M}{s}{p}{q}$ means
component $M$ guarantees the satisfaction of property $q$ up to and including
time $t + s \cdot dt$ under the assumption that property $p$ is satisfied at
time $t$, where $s$ is an update period of $M$, $t$ is the beginning of the
current update period, and $dt$ is the global tick.  
By induction over time, $\AG{M}{s}{true}{p}$ implies $M$ always satisfies $p$.

The choice of update period $s$ in $\AG{M}{s}{p}{q}$ is guided by the property $q$.
If $q$ is given in terms of the variables of a single-rate sub-component of $M$, then the 
update period of that sub-component is an appropriate choice for $s$.  
If $q$ involves variables from multiple sub-components, then the shortest update 
period of these sub-components is a good choice for $s$.

We use the following Assume-Guarantee rule as a formal proof rule for 
multi-rate systems because it is general enough to handle most cases of interest, 
including our case studies.  The rule we use is \emph{asymmetric} in the sense
that only one component makes assumptions about the other component.
Other A-G proof rules, including \emph{symmetric} ones (e.g.~\cite{barringer2003proof}),  
can also be used with our framework.
\vspace*{-.05in}
\begin{customrule}{AG}
\setlength{\jot}{0.1ex}
\setlength{\abovedisplayskip}{-5pt}
\label{rule:ag-rule}
	\begin{gather*}
	\AG{M_1}{s_1}{p}{q} \\
	\AG{M_2}{s_2}{true}{p} \\
	\rule[1pt]{3cm}{1pt} \\
	\AG{(M_1 \parallel M_2)}{s_1}{true}{q}
	\end{gather*}
\end{customrule}
\vspace*{-.05in}


Rule~\ref{rule:ag-rule} allows one to prove that a system composed of 
components $M_1$ and $M_2$ satisfies a property $q$ by proving that
(1)~component $M_1$ guarantees $q$ under assumption $p$, and 
(2)~component $M_2$ assures $p$ unconditionally.

\subsection{Coordinated Switching}
\label{sec:coordinated-switching}

\emph{Coordinated switching} is when a switch from AC to BC in one component forces
another component to also switch from AC to BC.
Suppose components $M_1$ and $M_2$ use Simplex to ensure their contracts 
$\scriptC[1]=(I_1, O_1, A_1, G_1)$ and 
$\scriptC[2]=(I_2, O_2, A_2, G_2)$, respectively.  Assuming $A_2 = true$ and 
$G_2 \models A_1$, where $\phi \models \varphi$ means $\phi$ entails $\varphi$, 
we have $M_1 \parallel M_2 \models G_1$ by applying
Rule~\ref{rule:ag-rule}. Let AC$_i$ and BC$_i$ be the AC and BC, respectively, of $M_i$. 
A typical situation that requires coordinated switching is as follows.
Suppose $G_2$ is the implication $\phi_1 \implies \phi_2$, and 
$\phi_1$ involves a shared variable that $M_1$ modifies when 
switching from AC$_1$ to BC$_1$ such that $\phi_1$ changes from $\textit{false}$ to
$\textit{true}$ as a result.  Suppose BC$_2$ ensures $\phi_2$ but AC$_2$ does not. When
$M_1$ uses AC$_1$, $G_2$ is vacuously true.  When $M_1$ uses BC$_1$, $M_2$ must use BC$_2$ to satisfy
$G_2$.  Thus, if $M_2$ is using AC$_2$ when $M_1$ switches to BC$_1$, then $M_2$ must perform a
coordinated switch to BC$_2$.

Coordinated switching is required in our case study.  
The \emph{Mission Planning} component has a BC called the recharge controller, which tells
the rover to go back to the most recently visited power station to recharge.  When
\emph{Mission Planning} switches to its BC, it assumes that the \emph{Navigation} component
can steer the rover to that power station within a certain energy budget.  
The BC in the \emph{Navigation} component guarantees this energy constraint,
but the AC does not.  Thus, when the \emph{Mission Planning}
component switches to its BC, the \emph{Navigation} component must also switch to 
its BC.  In this example, the shared variable in $\phi_1$ that communicates the need
for the \emph{Navigation} component to switch is the variable $ctlr$, which explicitly
indicates which controller is in control of the \emph{Mission Planning} component.

Coordinated switching can be generalized to be between \emph{modes}.  
A component may have multiple modes with different guarantees.
Each mode can use a Simplex instance to assure its guarantees.
When a component $M_1$ switches modes,
it may require component $M_2$ to switch to another mode so that
it can use the guarantee that $M_2$'s new mode provides.  
In our example above, we can view \emph{Mission Planning} as having 
two modes, named go-to-target and recharge, and \emph{Navigation} as having 
two modes, named go-to-target and backtrack.  
When \emph{Mission Planning} switches to recharge mode, it requires 
\emph{Navigation} to switch to backtrack mode.
In our case study, recharge mode and backtrack mode use certified controllers, so
we do not need to nest Simplex instances in them.

\section{The Quickbot 
Case Study}
\label{sec:case-study}
We conducted a detailed case study of our approach based on the Quickbot 
\cite{quickbot14} ground rover developed for the Coursera course Control 
of Mobile Robots (\url{https://www.coursera.org/course/conrob/}). The 
rover's mission is to visit a sequence of predetermined target locations.  
We consider a challenging version of the problem: the rover might need to 
recharge at power stations during the mission, and the rover 
does not have a map showing the locations of power stations or the locations 
and shapes of obstacles. 

\begin{figure}[htbp] 
	\begin{center} 
		\includegraphics[scale=0.34]{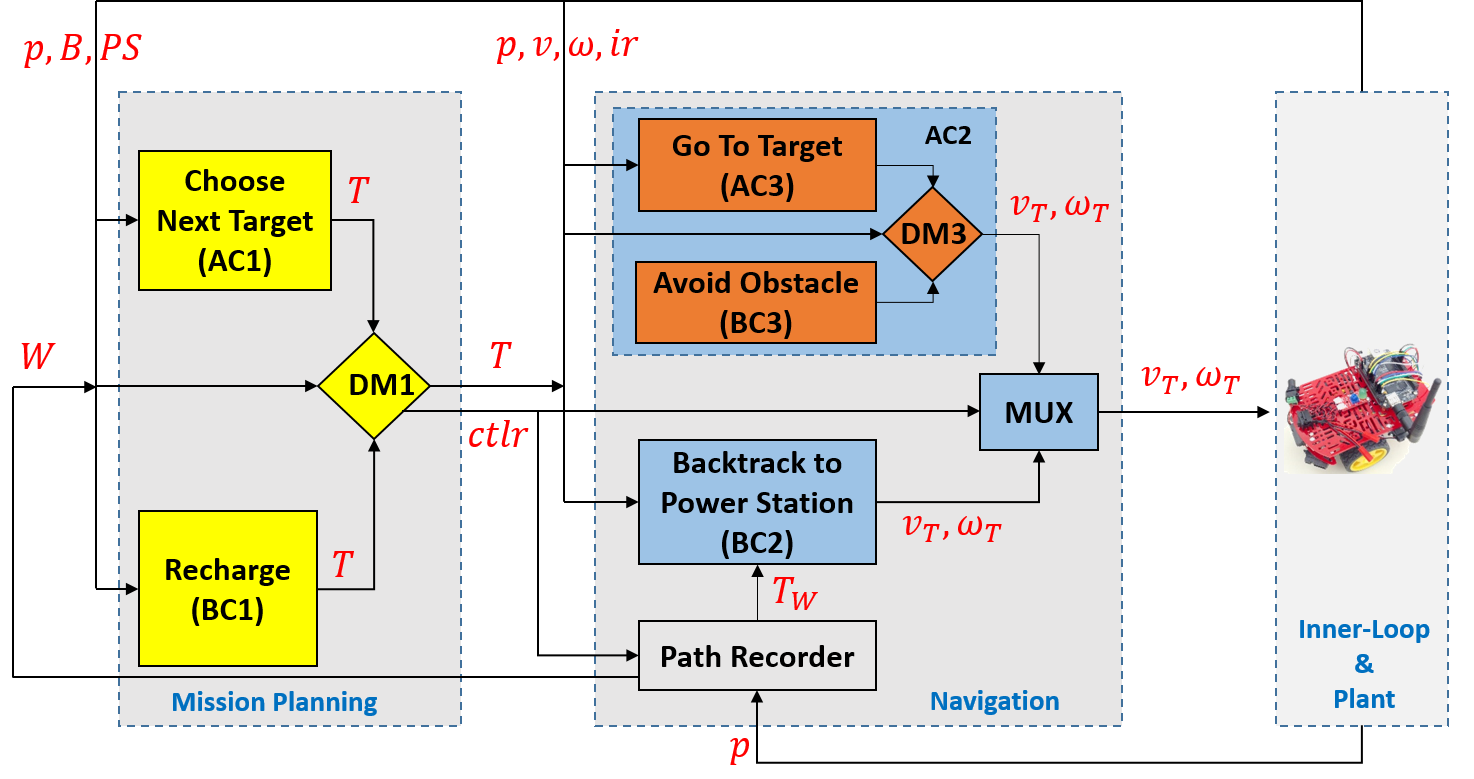} 
		\caption{Component-based Simplex architecture 
			for Quickbot rover. \cutoff{The $ctlr$ signal comes 
			from the \emph{Mission Planning} component's DM and is hardwired to 
			the multiplexer (MUX) in the \emph{Navigation} component so that 
			switching in \emph{Mission Planning} is cascaded to \emph{Navigation}, as 
			explained in Section~\ref{sec:quickbot-cbsa}.  The orange
            Simplex instance assures the CF property.  The yellow and 
            light-blue Simplex instances together assure the ES property. }}
		\label{fig:quickbot-simplex} 
	\end{center} 
    \vspace*{-0.2in}
\end{figure}

The top-level architecture for the Quickbot case study consists of three single-rate
components: \emph{Mission Planning}, \emph{Navigation}, \emph{Inner-Loop $\&$ Plant}; 
see Fig.~\ref{fig:quickbot-simplex}.  \emph{Mission Planning}'s update period $s_\MP$
is a multiple of \emph{Navigation}'s $s_\Nav$, which in turn is a multiple of \emph{Inner-Loop $\&$ Plant}'s $s_\Plant$.  The Quickbot system composed of these components is therefore a multi-rate component.
\emph{Mission Planning} sends the next target position $T$ to the \emph{Navigation} 
component.  \emph{Navigation} steers the rover to $T$ while 
avoiding obstacles.  It does this by computing an appropriate target linear
velocity vector ($v_T$) and target angular velocity ($\omega_T$) at each
time step.  The pair of velocities $(v_T, \omega_T)$ is sent 
to \emph{Inner-Loop $\&$ Plant}, which computes and actuates the appropriate 
rotational speeds for each of the two wheels to reach the desired 
velocities.

The rover is equipped with infrared (IR) sensors, two wheel
encoders (which count revolutions of the wheels), a power-station sensor, 
and a battery-level sensor.  \emph{Inner-Loop $\&$ Plant} converts the raw sensor data into 
the rover's current position ($p$), linear velocity vector ($v$), angular 
velocity ($\omega$), IR distances ($ir$) to obstacles, location of the last-detected power station ($PS$), and battery level ($B$). 

We use \MP, \Nav\ and \Plant\ to denote the \emph{Mission Planning}, \emph{Navigation} and \emph{Inner-Loop $\&$ Plant} components, respectively, and \QB\ to denote the entire Quickbot system.  \QB\ is the parallel composition of 
\MP, \Nav, and \Plant, i.e., $\QB = \MP \parallel \Nav \parallel \Plant$.  
\Plant\ is implicitly the parallel composition of the \emph{Inner-Loop Control} 
component and the physical plant.



\subsection{Problem Statement and Assumptions}
\label{sec:quickbot-problem}

We design a CBSA whose mission is to visit the specified 
targets while ensuring the following safety properties.
\begin{enumerate}
	\item \textit{Energy safety} (ES): the rover never runs out of 
	power.
	\item \textit{Collision freedom} (CF): the rover never collides with 
	an obstacle. 
\end{enumerate}

We make the following assumptions about the rover.
\begin{enumerate}
	\item The rover does not have a map of the environment.
    \item There are gaps (blind spots) between the fields-of-view of the distance sensors.  
	\item The distance sensors have a limited sensing range $[0, R_s]$.
    \item The rover can stop instantaneously.	
	\item The linear speed $v$ is bounded by $v \in [0, v_{max}]$.
 	\item The rotational speed $\omega$ is bounded by 
 	$\omega \in [-\omega_{max},  \omega_{max}]$.
    \item The rover can reach any pair of values $(v, \omega)$ instantaneously,
	provided they are within bounds.
    \item The rover starts at a power station.
	\item The rover can detect a power station when passing within 
	distance $d_{\PS}$ of it.
    \item The rover can recharge when it is within distance $d_{\PS}$ of a power station.
	\item The power consumption of the rover is a monotonically increasing 
    function of the rotational speeds of the two wheels.
\end{enumerate}

We make the following assumptions about the environment.
\begin{enumerate}
	\item Obstacles are stationary polyhedra.
	\item There is a known lower bound on the internal angles between 
	edges of an obstacle.
	\item There is a known lower bound on the edge lengths of an 
	obstacle.
	\item The separation between obstacles is such that whenever two 
	adjacent sensors simultaneously detect an obstacle, they are detecting the same 
	obstacle.
\end{enumerate}

Assumptions 1-7 about the rover and 1-4 about obstacles 
are the same as in~\cite{phan15collision}. 
This allows us to reuse the Simplex instance in~\cite{phan15collision}
to help ensure the CF property. 
Let $A_P$ denote the conjunction of the above assumptions about the rover. 
We assume \Plant\ satisfies $A_P$. 

\subsection{Quickbot CBSA}
\label{sec:quickbot-cbsa}

The architecture of the
Quickbot CBSA is shown in Fig.~\ref{fig:quickbot-simplex}.
A novelty of our case study is that we apply the Simplex architecture to 
the \emph{Mission Planning} component; Simplex is traditionally applied 
to lower-level controllers that interact 
directly with the plant. In our component-based architecture, 
\emph{Mission Planning} controls a \emph{virtual plant} comprising the 
\emph{Navigation} component, the \emph{Inner-Loop Control} component, and the physical plant.

{
\def\OldComma{,}
    \catcode`\,=13
    \def,{%
      \ifmmode%
        \OldComma\discretionary{}{}{}%
      \else%
        \OldComma%
      \fi%
    }%
\myparagraph{Quickbot Components} We illustrate Definition~\ref{def:component}
with the Quickbot components.
The single-rate \emph{Mission Planning} component is $\MP = (x_\MP, u_\MP, y_\MP, S_\MP)$, 
where
$x_\MP = \{T, ctlr\}$,
$u_\MP = \{p, B, \PS, W\}$,
$y_\MP = \{T, ctlr\}$,
and $S_\MP = (f_\MP, g_\MP, s_\MP)$.  Here $f_\MP$ and $g_\MP$ are the next-state and output functions representing the algorithm
for \MP described later in this section, and $s_\MP$ is the update period of \MP.
The behavior of \MP at global tick $i$ is specified by:
\begin{equation*}
\left\{
\begin{array}{lcl}
  x_\MP(i) = f_\MP\left(x_\MP(i-1), u_\MP(i)\right) \text{ if } i \text{ mod } s_\MP = 0\\
  y_\MP(i) = g_\MP\left(x_\MP(i), u_\MP(i)\right),  \text{ if } i \text{ mod } s_\MP = 0\\
\end{array}
\right.
\end{equation*}

The single-rate \emph{Navigation} component is $\Nav = (x_\Nav, u_\Nav, y_\Nav, S_\Nav)$,
where 
$x_\Nav = \{v_T, \omega_T, W\}$,
$u_\Nav = \{T,ctlr,p,v,\omega,ir\}$,
$y_\Nav = \{v_T, \omega_T,W\}$,
and $S_\Nav = (f_\Nav, g_\Nav, s_\Nav)$. Here $f_\Nav$ and $g_\Nav$ are the next-state and output functions representing the algorithm
for \Nav described later in this section, and $s_\Nav$ is the update period of \Nav. The behavior of \Nav at global tick $i$ is specified by:
\begin{equation*}
\left\{
\begin{array}{lcl}
  x_\Nav(i) = f_\Nav\left(x_\Nav(i-1), u_\Nav(i)\right) \text{ if } i \text{ mod } s_\Nav = 0\\
  y_\Nav(i) = g_\Nav\left(x_\Nav(i), u_\Nav(i)\right),  \text{ if } i \text{ mod } s_\Nav = 0\\
\end{array}
\right.
\end{equation*}

\noindent

The single-rate \emph{Inner-Loop $\&$ Plant} component is $\Plant = (x_\Plant, u_\Plant, y_\Plant, S_\Plant)$,
where 
$x_\Plant = \{p, v, \omega, ir, PS, B\}$,
$u_\Plant = \{v_T, \omega_T\}$,
$y_\Plant = \{p,v,\omega,ir,PS,B\}$,
and $S_\Plant = (f_\Plant, g_\Plant, s_\Plant)$. Here $f_\Plant$ and $g_\Plant$ are the next-state and output functions representing the
dynamics and sensing of \Plant, and $s_\Plant$ is the update period of \Plant. The behavior of \Plant at global tick $i$ is specified by:
\begin{equation*}
\left\{
\begin{array}{lcl}
  x_\Plant(i) = f_\Plant\left(x_\Plant(i-1), u_\Plant(i)\right) \text{ if } i \text{ mod } s_\Plant = 0\\
  y_\Plant(i) = g_\Plant\left(x_\Plant(i), u_\Plant(i)\right),  \text{ if } i \text{ mod } s_\Plant = 0\\
\end{array}
\right.
\end{equation*}

\noindent
where $f_\Plant$ and $g_\Plant$ are the next-state and output functions representing the
dynamics and sensing of \Plant.  The order of execution
is $\MP$ and then $\Nav$ (in time steps when they both execute) and then $\Plant$.  We assume 
$s_\MP$ is a multiple of $s_\Nav$, and $s_\Nav$ is a multiple of $s_\Plant$.
} 

\myparagraph{Quickbot Controllers}
Recall that we assume that the rover can
detect a power station within distance $d_\PS$, for example, by using 
a camera to read a QR code or by using an RFID reader to read a tag. 
When the rover passes near a power station, it can determine with IR 
sensors whether the power station is accessible from its current location;
i.e., there are no intervening obstacles.  If the power 
station is accessible, the rover remembers it as the last-visited (i.e.,
the most recently passed) power station.

The \emph{Mission Planning} AC informs the \emph{Navigation} component
of the next target to be visited to fulfill the mission goals, 
while the BC (the recharge controller) instructs the \emph{Navigation} component 
to navigate to the most recently detected power station.
We say that the rover is in {\em recharge mode} when DM$_1$ has switched control 
to the recharge controller.
To guarantee that recharge mode will drive the rover 
to a power station without depleting the battery, we need a certified
backtrack-to-power-station controller in the \emph{Navigation} component. 
Since the rover lacks a map of the environment,
backtracking is the most realistic way to guarantee that the rover can 
reach a power station without running out of power.

We also briefly considered a variant of the problem in which the rover 
has a map showing locations of power stations. If the rover does not also
have a map of obstacles, backtracking to the last-visited 
power station might still be necessary, because nearby power stations could be 
blocked by unknown obstacles that would take too much power to circumnavigate.  
More complex strategies for the AC in the \emph{Mission Planning} component could 
help avoid backtracking.  For example, instead of going as 
directly as possible to the targets, it might add waypoints that are 
slightly out of the way but help the rover determine accessibility of 
nearby power stations.

Note that there is coordinated switching between the \emph{Mission Planning} 
component and the \emph{Navigation} component. When \emph{Mission Planning} 
switches to the recharge controller, \emph{Navigation} must switch to the backtrack controller.  
The recharge controller in \emph{Mission Planning} assumes 
that the \emph{Navigation} component can navigate the rover to a power station 
using at most a specified amount of energy.  The go-to-target controller 
does not provide such guarantees, whereas the backtrack-to-power-station 
controller (backtrack controller, for short) does. 
This cascading switch is implemented by hardwiring a decision signal 
from \emph{Mission Planning}'s DM to  \emph{Navigation}'s DM,
as shown in Fig.~\ref{fig:quickbot-simplex}. 

We introduce a recorder module that records the positions and heading angles 
(waypoints) at every time step of \emph{Navigation}.  When in recharge mode, the 
backtrack controller tracks these recorded positions in reverse 
order.  We also want the backtrack controller to preserve the \CF\ property
so that the rover is guaranteed not to collide with any static obstacles
on the way back to the last-visited power station.  
\extendedonly{ 
We consider three
designs for the backtrack controller.

\begin{enumerate}
	\item Instead of recording waypoints, we can record the control inputs
	$(v, \omega)$ for \Plant\ in every \emph{Navigation} time step.  When
	backtracking, we replay $(v, -\omega)$.  This controller will trace
	the forward path exactly provided that there are no actuation or state 
	estimation errors.  Otherwise, the error will
    accumulate.
	\item The rover can backtrack from its current position to a 
	recorded waypoint by first turning in place to point to the target
	waypoint and then moving in a straight line to it.  This 
	approach is robust, but it is slow, consumes more energy, and replaces a curve in the 
    forward trajectory with a line segment in the backward trajectory.  Such a discrepancy
    could lead to collisions unless the avoid-obstacles controller is invoked.
\item The rover can backtrack from its current position to a 
	recorded waypoint by computing the appropriate values of 
	$(v, \omega)$.  This means we need to solve the \emph{inverse kinematics
	problem}, which in general has no analytical solution.  We can, however, obtain the least-squares solution, which has several benefits.
	If there are no errors then the least-squares solution is an 
	exact solution; i.e., this approach falls back to the first one.
	If there are errors, the errors will be minimized in the 
	least-squares sense and bounded, instead of accumulating.  Assuming the errors are small, 
    the trajectory during backtracking is always close to the forward trajectory, 
    and the energy consumed during backtracking is within a bounded margin
	of the energy consumed going forward.  We adopt this approach for the backtrack
	controller because of these benefits.
\end{enumerate}
} 
The rover can backtrack from its current position to a recorded waypoint by 
computing the appropriate values of $(v, \omega)$.  It can do this by
solving the \emph{inverse kinematics problem} for the least-squares solution.
Suppose the rover is currently at point $A = (x_1, y_1, \theta_1)$ 
and wants to reach a recorded waypoint $B = (x_2, y_2, \theta_2)$ in one 
Navigation time step $t_\Nav$.  The kinematics equation is given below.
\begin{equation}
\left\{
\setlength\arraycolsep{4pt}
\begin{array}{lcl}
x_2 & = & x_1 + \int_{t_1}^{t_1+t_\Nav} v\cos(\theta_1 + \omega t)dt \\
y_2 & = & y_1 +  \int_{t_1}^{t_1+t_\Nav} v\sin(\theta_1 + \omega t)dt \\
\theta_2 & = & \theta_1 + \omega t_\Nav
\end{array}
\right.
\label{eq-kinematic}
\end{equation}
The solution in the trivial case where $\theta_1 = \theta_2$ 
is $\omega = 0$ and $v = {\it AB}/t_\Nav$.  If $\theta_1 \neq \theta_2$,
Eq.~\ref{eq-kinematic} becomes:
\begin{equation}
\left\{
\setlength\arraycolsep{4pt}
\begin{array}{lcl}
x_2 & = & x_1 + \frac{v}{\omega}(\sin\theta_2 - \sin\theta_1) \\
y_2 & = & y_1 + \frac{v}{\omega}(\cos\theta_1 - \cos\theta_2) \\
\theta_2 & = & \theta_1 + \omega t_\Nav
\end{array}
\right.
\label{eq-kinematic-simplified}
\end{equation}

We solve Eq.~\ref{eq-kinematic-simplified} for the least-squares solution
using the standard least-squares method.

We reuse the Simplex instance in~\cite{phan15collision} 
to help ensure the CF property.  We nest this instance inside the 
AC of the Simplex instance in \emph{Navigation}.
The intuition behind this nested composition of Simplex instances is 
that the backtrack controller also helps ensure the CF property,
because it re-traces a collision-free path the rover has already traveled.
Therefore, the Simplex instance involving AC$_2$ and BC$_2$ in Fig.~\ref{fig:quickbot-simplex} assures the CF property.

\myparagraph{Contracts of Quickbot Components}
Based on the above analysis, we specify the following contracts.
The \emph{Mission Planning} component's contract is:
\begin{quote}
Inputs: $p, \PS, B, W$ \\
Outputs: $T, ctlr$ \\
Assumption: $ctlr = \text{BC} \implies$ \\
\hspace*{.7in} $\BE(p, \PS, W) \le (1 + \epsilon_\BE)\FE(\PS, p)$ \hfill \\
Guarantee: $B > E(p, \PS)$
\end{quote}
\noindent
where $\BE(p, \PS, W)$ is the energy needed to backtrack from the current
position $p$ to the most recently visited power station $\PS$ going
through the sequence of recorded waypoints $W$; $\epsilon_\BE$
is a constant defined in Section~\ref{sec:es-switching-logic};
$\FE(\PS, p)$ is the
energy expended on the forward path from $\PS$ to $p$; $E(p, \PS)$ 
is the energy needed to go from $p$ to $\PS$, including the amount of
energy needed to turn 180\degree\ before backtracking.  The \emph{Navigation} component's contract is:
\begin{quote}
Inputs: $T, ctlr, p, ir, v, \omega$ \\
Outputs: $v_T, \omega_T, W$ \\
Assumption: $A_P$ \\
Guarantee: $(d_o > 0) \wedge (ctlr = \text{BC} \implies$ \hfill \\
\hspace*{.7in} $\BE(p, \PS, W) \le (1 + \epsilon_\BE)\FE(\PS, p))$
\end{quote}
\noindent
where $d_o$ is the shortest distance to an obstacle.
The \emph{Inner-Loop $\&$ Plant}'s contract is:
\begin{quote}
Inputs: $v_T, \omega_T$ \\
Outputs: $p, \PS, B, ir, v, \omega$ \\
Assumption: $true$ \\
Guarantee: $A_P$
\end{quote}
The input and output variables of each component are shown in
Fig.~\ref{fig:quickbot-simplex}.

\subsection{Switching Logic for the ES Property}
\label{sec:es-switching-logic}

The ES property can be formally expressed as $G(B > 0)$, where $G$ is 
the \emph{always} (global) operator in linear temporal logic.
In our design, this is ensured by having the rover recharge at a power 
station whenever the battery level is low.  So, we enforce the stricter property that 
the rover always has enough energy to return to the last-visited power station.  
This property is formalized as $G(B > E(p, \PS))$, where $E(p, \PS)$ is the amount 
of energy needed to go from the current position $p$ to the last-visited power station $\PS$, including the amount of energy needed to 
turn 180\degree\ ($E_{180}$) before going back.  Hereafter, we refer to 
$G(B > E(p, \PS))$ as the ES property.

We assume the following constants are known:
\begin{enumerate}
	\item $\epsilon_\BE$: a constant such that the
    backtracking energy \BE(p, \PS, W) needed to backtrack
    from the current position $p$ to the last-visited power station \PS\ 
    through a sequence of recorded waypoints $W$ is bounded by $(1 + \epsilon_\BE)$ 
    times the forward energy \FE(\PS, p) expended since $PS$ was last visited, i.e.,
    $\BE(p, \PS, W) \le (1 + \epsilon_\BE)\FE(\PS, p)$.  If there are no errors in actuation or state estimation, we can take $\epsilon_\BE=0$.
	\item $E_\MP$: the worst-case energy expended in one update period of \emph{Mission Planning} by an arbitrary controller.
    \item $E_{180}$: the energy needed to turn in place by 180\degree.
	\item $\BE_\MP$: a bound on the energy needed to backtrack from the rover's position at the end
    of the next update period of \emph{Mission Planning} to the current position.
\end{enumerate}
Let $E_\MP$ be the amount of energy expended when both wheels rotate at 
maximum rotational speed for one update period of \emph{Mission Planning}.  $\BE_\MP$ equals $E_\MP$.

The DM in the \emph{Mission Planning} component decides 
whether to continue to go forward or to return to \PS.  If it
decides to go forward, it must ensure that the ES property holds at 
its next decision point.  The worst-case amount of energy needed to 
travel to a position within one update period and then backtrack from 
that position to \PS\ is 
$E_\MP + E_{180} + \BE_\MP + (1 + \epsilon_\BE)\FE(\PS, p)$.  
The switching condition is thus:
\begin{equation}
B \le E_\MP + E_{180} + \BE_\MP + (1 + \epsilon_\BE)\FE(\PS, p)
\label{eq:mp-switching-condition}
\end{equation}

This switching condition is derived compositionally by 
relying on \emph{Mission Planning}'s assumption, an assumption
that is later discharged by composing \emph{Mission Planning} with \emph{Navigation}.  
We do not need to consider implementation details of other components 
in the derivation process.



\subsection{Proof Outlines of ES and CF}
\label{sec:proof-outline}

\begin{lemma}
	\label{lem:1}
For any property $\phi$,
$\AG{\Nav}{s_\Nav}{A_P}{\phi} \implies  \AG{(\Nav \parallel \Plant)}{s_\Nav}{true}{\phi}$.
\end{lemma}
\vspace*{-0.2in}
\begin{proof}
	The proof follows from our assumptions and one application of Rule~\ref{rule:ag-rule}.
    \begin{gather*}
		\AG{\Nav}{s_\Nav}{A_P}{\phi} \\
		\AG{\Plant}{s_\Plant}{true}{A_P} \\
		\rule{4cm}{1pt} \\
		\AG{\Nav \parallel \Plant}{s_\Nav}{true}{\phi}
	\end{gather*}
\end{proof}

\begin{theorem}
	\label{thm:2}
	$\AG{\QB}{s_\MP}{true}{\ES}$
\end{theorem}
\myparagraph{Proof outline}
\begin{enumerate}
	\item We prove $\AG{\MP}{s_\MP}{A_{\BE}}{\ES}$, 
	where $A_{\BE}$ is the assumption in $\MP$'s contract given in Section~\ref{sec:quickbot-cbsa}.
	\item We prove $\AG{\Nav}{s_\Nav}{A_P}{A_{\BE}}$.  By Lemma~\ref{lem:1}, this implies
	$\AG{(\Nav \parallel \Plant)}{s_\Nav}{true}{A_{\BE}}$. 
	\item Applying Rule~\ref{rule:ag-rule}, we conclude that \\
	$\AG{(\MP \parallel \Nav \parallel \Plant)}{s_\MP}{true}{\ES}$.
\end{enumerate}

\begin{theorem}
	$\AG{\QB}{s_\MP}{true}{\CF}$
\end{theorem}
\myparagraph{Proof outline}
\begin{enumerate}
	\item We prove $\AG{\Nav}{s_\Nav}{A_P}{\CF}$. 
    By Lemma~\ref{lem:1}, this implies
	$\AG{(\Nav \parallel \Plant)}{s_\Nav}{true}{\CF}$.
	\item Applying Rule~\ref{rule:ag-rule}, we conclude that \\ 
	$\AG{(\MP \parallel \Nav \parallel \Plant)}{s_\MP}{true}{\CF}$.
\end{enumerate}

\shortonly{
The proofs of the ES and CF properties are provided in an extended version
of this paper~\cite{phan2017cbsa}.
}

\extendedonly{
The full proofs of the ES and CF properties are provided in Appendix~\ref{appendix-a}.
}

\subsection{Experimental Results}
\label{sec:experiments}

We implemented the CBSA for the Quickbot rover in Matlab using
the following parameter values: 
(1)~number of distance sensors $N = 8$; 
(2)~angle of detection of the sensors $\beta_s = 5^{\circ}$; 
(3)~maximum range of the sensors $R_s = 0.8 m$;
(7)~radius of the wheels $r = 0.0325 m$;
(8)~distance between the centers of the two wheels $l = 0.09925 m$;
(9)~maximum linear velocity $v_{max} = 0.8 m/s$;
(10)~maximum angular velocity $\omega_{max} = 7\pi\ rad/s$;
(11)~power station sensor detection range $d_\PS = 0.1 m$;
(12)~maximum battery level $B_{max} = 100$.
We use a power model that is an affine function of the angular velocities of
the two wheels: $P(\omega_l, \omega_r) = p_1(\abs{\omega_l} + \abs{\omega_r}) + p_2$, 
where $p_1$ and $p_2$ are constants and $\omega_l, \omega_r$
are the rotational speeds of the two wheels. 
\extendedonly{These are calculated from
linear velocity $v$ and angular
velocity $\omega$ as follows.
\begin{equation}
\left\{
\begin{array}{lcl}
\omega_l & = & \frac{2v - \omega l}{2r} \\
\omega_r & = & \frac{2v + \omega l}{2r}
\end{array}
\right.
\end{equation}
} 
In our experiments, we use $p_1 = 0.15, p_2 = 0.01$.  The constants used
in Eq.~\ref{eq:mp-switching-condition} are $E_\MP = \BE_\MP = 2.032$, and
$E_{180} = 1.524$, which are computed based on $v_{max}$ and $\omega_{max}$. We also choose $\epsilon_\BE = 0$ as we assume there are no errors.  The global tick is $dt = 0.05 s$.
The update periods of \emph{Mission Planning}, \emph{Navigation}, and \emph{Inner-Loop $\&$ Plant} 
are $4dt, 2dt$ and $dt$, respectively.
We adopt the algorithm from the 
Coursera course ``Control of Mobile Robots'' for the go-to-target controller.
The avoid-obstacles controller simply stops the rover, as in~\cite{phan15collision}.
The backtrack controller implements the least-squares approach described
in Section~\ref{sec:quickbot-cbsa}.  In \emph{Mission Planning},
the choose-next-target controller picks the next target in
a sequence of pre-determined target locations when the rover arrives at the
previous target.  The recharge controller simply sets $T = \PS$.

\begin{figure}[htbp] 
	\begin{center} 
		\includegraphics[scale=0.38]{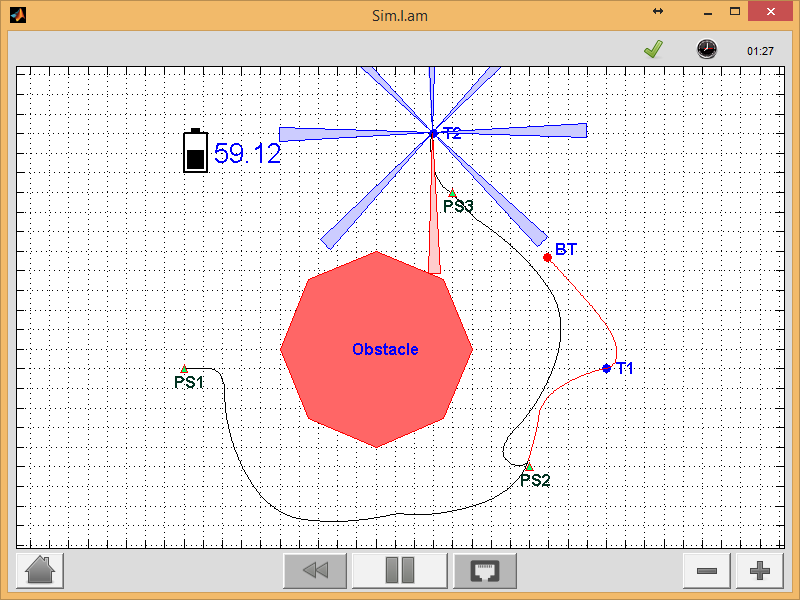} 
		\caption{Snapshot from a simulation showing the trajectory the rover
        takes to visit targets $T_1$ and $T_2$ while ensuring the ES and CF properties.}
		\label{fig:experiment-1} 
	\end{center} 
\end{figure}

Fig.~\ref{fig:experiment-1} shows the complete trajectory the rover takes with the following
configuration.  The rover's starting position is $P_0 = \PS_1 = (-1, 0, 0)$, its initial heading angle is
$\theta_0 = 0$, and the targets are $T_1 = (1.2, 0)$ and $T_2 = (0.3, 1.2)$, which must be visited in that order.
The power stations are located at $\PS_1 = (-1, 0), \PS_2 = (0.8, -0.5), \PS_3 = (0.4, 0.9)$.
The black lines represent the forward paths and the red line represents the backward path.  
The rover is able to reach $T_1$ without having to recharge.  At position 
$\BT = (0.895, 0.570)$ on its way from $T_1$ to $T_2$, the battery level drops to 
$B = 29.07$, triggering recharge mode.  The rover re-traces the forward path
segment $\PS_2 \BT$ exactly (the red and black lines are indistinguishable)
to recharge at power station $\PS_2$.  The battery level when it reaches $\PS_2$ is $B = 0.07$, indicating a tight switching condition.
After recharging to full capacity at $\PS_2$, the rover takes a different path to 
the last target $T_2$.  
A video of the simulation can be viewed at \url{https://youtu.be/i8WGVD5Vk7U}.

\section{Quickbot Mission Completion}
\label{sec:MC}

The Simplex architecture is not intended for assuring general liveness properties.  This is reflected by the fact that a Decision Module looks ahead one update period to determine if the (safety)
property in question is about to be violated; as such, it will not be able to detect a violation of a liveness property if the property's time horizon is unbounded.  As we show here, however, the Simplex architecture can be made to work with \emph{bounded} liveness properties.  In particular, we design a CBSA for the Quickbot that assures a property called \emph{Mission Completion} (MC).

Recall that the mission of the Quickbot rover is to visit a sequence of predetermined targets.  The MC property ensures that the rover
completes its mission within a given amount of time.  This property is interesting in the context of Simplex not only because it is a (bounded) liveness property, but also because it is a higher-level mission-oriented property, several layers of control removed from the physical plant.  The only physical parameter of the plant in question is (real) time.  The MC property is also more software-oriented in nature, as it revolves around a data structure (a list of mission targets).  The MC property can be formally expressed as $G(F_{<T}(\text{all targets are visited}))$, where $G$ and $F$ are the \emph{always} and \emph{eventually} operators in
linear temporal logic, and $T$ is the upper bound on amount of time the rover needs to complete its mission.
The unbounded version of the MC property is $G(F(\text{all targets are visited}))$. 
 In other words, it is always the case that the rover will eventually visits all targets.

In designing the CBSA for the MC property, we assume that obstacles are stationary and there is a \emph{map} showing 
the targets and the location and shape of every obstacle.  
For simplicity, we also assume that the rover has enough battery to operate in the given mission-completion timeframe.

Similar to the case study in Section~\ref{sec:case-study}, the CBSA consists of three single-rate components: \emph{Mission Planning}, \emph{Navigation}, and \emph{Inner-Loop $\&$ Plant}.
The main objective of the \emph{Mission Planning} component is to inform \emph{Navigation} where the next target is.  The BC for \emph{Mission Planning} simply outputs one target after another, whereas the AC can generate intermediate \emph{waypoints} 
in between targets.  The \emph{Mission Planning} component does not output the next waypoint or target until the
rover arrives at the current waypoint/target.

The AC can choose waypoints to 
optimize some criteria according to its path-planning algorithm.  Since the algorithm used by AC can be uncertified, it may generate a time-consuming path that violates the MC 
property.  Another factor that may cause a violation of MC is the manner in which \emph{Navigation}'s AC drives the rover through the waypoints to get to targets.

The \emph{Navigation} component steers the rover to the target or waypoint computed by 
the \emph{Mission Planning} component.  As such, it plays an important role in assuring the MC
property. \emph{Navigation} alone, however, cannot guarantee MC because it does not know the sequence of targets.
What it can do is to guarantee a bound on the time needed to move from one location to another
on the map when \emph{Mission Planning} activates its BC.  The BC in \emph{Navigation} is
designed to guarantee this bound.

The DM in \emph{Mission Planning} uses the bound to 
check whether a violation of MC is imminent.  We say the rover is about to violate the MC property if, in the next time step, it may end up at a location from which the \emph{Navigation}'s BC is unable to navigate through the remaining targets 
in time.  When \emph{Mission Planning}'s DM decides to switch, the switching is cascaded to \emph{Navigation}.

The contracts for \emph{Mission Planning} and \emph{Navigation} are as follows.
The \emph{Mission Planning} component's contract is:
\begin{quote}
Inputs: $p$ \\
Outputs: $T, ctlr$ \\
Assumption: $ctlr = \text{BC} \implies$ \\  
\hspace*{.7in} $\forall (p_1, p_2), t(p_1, p_2) \le tu(p_1, p_2) $ \\
Guarantee: $t(p, Tseq) \le \it{remaining\_time}$
\end{quote}
\noindent
where $p$ is the current position of the rover, $T$ is the output waypoint or target,
$ctlr \in \{\text{AC}, \text{BC}\}$ indicates which controller is active in
\emph{Mission Planning}, $t(p_1, p_2)$ is the actual time needed to go from location
$p_1$ to $p_2$, $tu(p_1, p_2)$ is the upper bound on $t(p_1, p_2)$ and is based on
\emph{Navigation}'s BC, $Tseq$ is the sequence of remaining targets, 
$t(p, Tseq)$ is the actual time needed to complete the mission from $p$, i.e.,
$t(p, Tseq) = t(p, Tseq[1]) + \sum_{i=1}^{n-1}{t(Tseq[i], Tseq[i+1])}$ where $n$ is the size
of $Tseq$, and $\it{remaining\_time}$ is the time until the mission completion deadline.
The \emph{Navigation} component's contract is:
\begin{quote}
Inputs: $T, ctlr, p, ir, v, \omega$ \\
Outputs: $v_T, \omega_T$ \\
Assumption: $A_P$ \\
Guarantee: $ctlr = \text{BC} \implies$ \\ 
\hspace*{.7in} $\forall (p_1, p_2), t(p_1, p_2) \le tu(p_1, p_2)$
\end{quote}
\noindent
where $ir, v, \omega, v_T$, and $\omega_T$ are as described in Section~\ref{sec:case-study},
$A_P$ is the conjunction of assumptions~2-7 about the rover given in Section~\ref{sec:case-study}.

The \emph{Inner-Loop $\&$ Plant}'s contract is similar to the one in Section~\ref{sec:case-study}.
For the actual implementation of \emph{Navigation}'s BC, we can use the A* algorithm~\cite{hart1968formal} to 
generate a collision-free path between two points on the map.  To bound the time 
$tu(p_1, p_2)$ needed to traverse this path, the BC can rely on just two motion primitives: turning in-place and moving 
straight ahead at a constant speed.
The \emph{Mission Planning}'s DM needs a bound on the time needed for \emph{Navigation}'s
BC to complete the entire mission.  It can use
$tu(p_1, p_2)$ for this bound, which can be calculated on-the-fly.  If the map is not too large, we can divide the map
into grid cells and pre-compute the maximum time needed for mission completion from each
cell.

The switching condition is then derived as follows.  We first compute the
region reachable by the rover in one update period of \emph{Mission Planning}.  
We then compute the maximal $t(p', Tseq)$, for $p'$ in the reachable region.
The switching condition is then $t(p', Tseq) < \it{remaining\_time} - s_\MP dt$,
where $s_\MP dt$ is the update period of \emph{Mission Planning}.  
We now provide a proof outline of the MC property.

\begin{theorem}
	\label{thm:4}
	$\AG{\QB}{s_\MP}{true}{\MC}$
\end{theorem}
\myparagraph{Proof outline}
\begin{enumerate}
	\item We prove $\AG{\MP}{s_\MP}{A_{\TU}}{\MC}$, 
	where $A_{\TU}$ is the assumption in $\MP$'s contract given above.
	\item We prove $\AG{\Nav}{s_\Nav}{A_P}{A_{\TU}}$.  By Lemma~\ref{lem:1}, this implies
	$\AG{(\Nav \parallel \Plant)}{s_\Nav}{true}{A_{\TU}}$. 
	\item Applying Rule~\ref{rule:ag-rule}, we conclude that \\
	$\AG{(\MP \parallel \Nav \parallel \Plant)}{s_\MP}{true}{\MC}$.
\end{enumerate}



\section{Conclusions}
\label{sec:conclusion}

We have presented a component-based Simplex architecture for
assuring the runtime safety of component-based cyber-physical systems,
and a detailed case study that illustrates how our proposed CBSA
helps the Quickbot ground rover assure two safety properties: 
energy safety and collision freedom.  We also presented a CBSA for
Quickbot that ensures a bounded liveness property called mission completion.
As future work, we plan to extend our case study to accommodate uncertainties in
actuation and sensor readings, and to investigate the application 
of our component-based Simplex architecture to UAVs and squadrons of UAVs. 
Another direction for future work is to develop a design process for 
CBSA that can be used to determine what guarantees each 
component should provide to ensure a global property.\\


\section*{Acknowledgment}
This work is supported in part by %
AFOSR Grant FA9550-14-1-0261, 
NSF Grants
CNS-1421893, 
and CCF-1414078, 
and ONR Grant N00014-15-1-2208. 
Any opinions, findings, and conclusions or recommendations expressed in
this material are those of the author(s) and do not necessarily reflect
the views of these organizations.

\bibliographystyle{IEEEtran}
\bibliography{cbsa}

\extendedonly{
  \appendix
  \section{Proof of ES and CF Properties}
\label{appendix-a}

The proofs below follow the proof outlines given in Section~\ref{sec:proof-outline}.

\subsection{Proof of ES property}
\label{sec:app-es-proof}



\myparagraph{Proof of  $\AG{\MP}{s_\MP}{A_{\BE}}{\ES}$}
We prove $\AG{\MP}{s_\MP}{A_{\BE}}{\ES}$ by showing that if the ES property
holds at the end of time step $i$,
then it holds at the end of time step $i + s_\MP$, i.e., the next period of \MP,
where $i \text{ mod } s_\MP = 0$.  The proof is easily extended to show that the ES property holds continuously during the time step.
Let $B', p', W', ctlr'$ denote the values of $B, p, W, ctlr$, respectively,
at the end of time step $i + s_\MP$.  We prove that if $B > E(p, PS)$ then
$B' > E(p', PS)$.  Since the rover
traces through a sequence of recorded waypoints $W$ when going back to
$PS$, we have 
\begin{equation}
E(p, PS) = \left\{
	\begin{array}{lcl}
		E_{180} + \BE(p, PS, W) \text{, if } ctlr = \text{AC} \\
        \BE(p, PS, W) \text{, if } ctlr = \text{BC}
    \end{array}
\right.
\label{eq:e_p_PS}
\end{equation}
There are four cases:

\myparagraph{Case 1:} $ctlr = \text{BC}$ and $ctrl' = \text{BC}$.
The rover is in recharge mode.  The backtracking energy needed
to backtrack from current position $p$ to the next position $p'$ is
$\BE(p, p', W)$.  Therefore the battery level at the end of the next decision period 
of \MP\ is
\begin{equation}
	B' = B - \BE(p, p', W) 
    \label{eq:case-1-1}
\end{equation}
We assume that the ES property holds at current time, i.e.,
\begin{equation}
	B > \BE(p, \PS, W)
\label{eq:case-1-2}
\end{equation}
We want to prove that the ES property still holds at end of the next
decision period of \MP\, i.e.,
\begin{equation}
	B' > \BE(p', \PS, W')
\label{eq:case-1-3}
\end{equation}
The backtracking energy from current position $p$ to $PS$ can be
partitioned into $\BE(p, p', W)$ and $\BE(p', \PS, W')$, i.e.,
\begin{equation}
	\BE(p, \PS, W) = \BE(p, p', W) + \BE(p', \PS, W')
    \label{eq:case-1-4}
\end{equation}
Re-arranging this equation gives:
\begin{equation}
	\BE(p', \PS, W') = \BE(p, \PS, W) - \BE(p, p', W)
    \label{eq:case-1-5}
\end{equation}
Combining Eq.~\ref{eq:case-1-2} and Eq.~\ref{eq:case-1-5}, we have:
\begin{equation}
	\BE(p', \PS, W') < B - \BE(p, p', W)
    \label{eq:case-1-6}
\end{equation}
Combining Eq.~\ref{eq:case-1-1} and Eq.~\ref{eq:case-1-6}, we have:
\begin{equation}
	\BE(p', \PS, W') < B'
    \label{eq:case-1-7}
\end{equation}
According to Eq.~\ref{eq:e_p_PS}, $E(p', PS) = \BE(p', \PS, W')$, therefore:
\begin{equation}
	B' > E(p', PS). 
\end{equation}

\myparagraph{Case 2:} $ctlr = \text{AC}$ and $ctrl' = \text{AC}$.
The energy needed to go from current position $p$ to
the next position $p'$ is $\FE(p, p') \le E_\MP$.
The battery level at the end of the next decision period of \MP\ is:
\begin{equation}
	B' \ge B - E_\MP
\label{eq:case-2-1}
\end{equation}
We assume that the ES property holds at the current time, i.e.,
\begin{equation}
	B > E_{180} + \BE(p, \PS, W)
\label{eq:case-2-2}
\end{equation}
We want to prove that the ES property still holds at the end of the next
decision period of \MP\, i.e.,
\begin{equation}
	B' > E_{180} + \BE(p', \PS, W')
\label{eq:case-2-3}
\end{equation}
Note that if the rover detects a new power station $\PS'$ on the way
from $p$ to $p'$ then the ES property still holds at the end of the
next decision period, because $\BE(p', \PS', W') \le \BE(p', \PS, W')$.
$ctrl = \text{AC}$ means the switching condition is false, i.e.,
\begin{equation}
	B > E_\MP + E_{180} + \BE_\MP + (1 + \epsilon_\BE)\FE(\PS, p)
\label{eq:case-2-4}
\end{equation}
Since $\BE(p, \PS, W) \le (1 + \epsilon_\BE)\FE(\PS, p)$, we have:
\begin{equation}
	B > E_\MP + E_{180} + \BE_\MP + \BE(p, \PS, W)
\label{eq:case-2-5}
\end{equation}
Combining Eq.~\ref{eq:case-2-1} with Eq.~\ref{eq:case-2-5}, we have:
\begin{equation}
	B' > E_{180} + \BE_\MP + \BE(p, \PS, W)
\label{eq:case-2-6}
\end{equation}
$\BE(p', \PS, W')$ can be partitioned into $\BE(p', p, W')$
and $\BE(p, \PS, W)$, i.e.,
\begin{equation}
	\BE(p', \PS, W') = \BE(p', p, W') + \BE(p, \PS, W)
\label{eq:case-2-7}
\end{equation}
Energy needed to backtrack from $p'$ to $p$ is bounded by $\BE_\MP$,
thus:
\begin{equation}
	\BE(p', \PS, W') \le \BE_\MP + \BE(p, \PS, W)
\label{eq:case-2-8}
\end{equation}
Combining Eq.~\ref{eq:case-2-6} with Eq.~\ref{eq:case-2-8}, we have:
\begin{equation}
	B' > E_{180} + \BE(p', \PS, W')
\label{eq:case-2-9}
\end{equation}
According to Eq.~\ref{eq:e_p_PS}, $E(p', \PS) = E_{180} + \BE(p', \PS, W')$, 
therefore,
\begin{equation}
	B' > E(p', \PS).
\label{eq:case-2-10}
\end{equation}

\myparagraph{Case 3:} $ctlr = \text{AC}$ and $ctrl' = \text{BC}$.
The rover is switching from AC to BC, which means it must turn 180\degree\
before backtracking.  The battery level in the next decision period of \MP\ is:
\begin{equation}
	B' = B - E_{180} - \BE(p, p', W)
\label{eq:case-3-1}
\end{equation}
We assume that the ES property holds at the current time, i.e.,
\begin{equation}
	B > E_{180} + \BE(p, \PS, W)
\label{eq:case-3-2}
\end{equation}
We want to prove that the ES property still holds at the end of the next
decision period of \MP\, i.e.,
\begin{equation}
	B' > \BE(p', \PS, W')
\label{eq:case-3-3}
\end{equation}
Observe that if we let $B_{180} = B - E_{180}$ then Case~3 becomes
Case~1 with the same proof.

\myparagraph{Case 4:} $ctlr = \text{BC and } ctrl' = \text{AC}$.
When \MP\ switches from BC to AC, the switching condition must 
evaluate to false, i.e.,
\begin{equation}
	B > E_\MP + E_{180} + \BE_\MP + (1 + \epsilon_\BE)\FE(\PS, p)
\label{eq:case-4-1}
\end{equation}
Thus, this case folds back to Case~2 with the same proof.

\myparagraph{Proof of $\AG{\Nav}{s_\Nav}{A_P}{A_{\BE}}$}

Since the decision $ctlr$ from \MP\ is hardwired 
to the switch in \Nav, we only need to prove that the backtrack
controller in \Nav\ satisfies $\BE < (1 + \epsilon_\BE)\FE(\PS, p)$.
As discussed in Section~\ref{sec:quickbot-cbsa}, the sequence of
values of $(v, \omega)$ on the backtrack
path will have the same magnitudes as the ones on the forward path.
Therefore the backtrack energy will be the same as the forward energy.

\subsection{Proof of CF Property}
\label{sec:app-cf-proof}


\myparagraph{Proof of $\AG{\Nav}{s_\Nav}{A_P}{\CF}$}

{
\def\OldComma{,}
    \catcode`\,=13
    \def,{%
      \ifmmode%
        \OldComma\discretionary{}{}{}%
      \else%
        \OldComma%
      \fi%
    }%
We prove $\AG{\Nav}{s_\Nav}{A_P}{\CF}$ 
by proving that\\ $\AG{AC2}{s_\Nav}{A_P}{\CF}$ and $\AG{BC2}{s_\Nav}{A_P}{\CF}$.  
Since AC2 is the same Simplex instance 
in~\cite{phan15collision}, the proof of \\ $\AG{AC2}{s_\Nav}{A_P}{\CF}$ 
is already available there.  We show that $\AG{BC2}{s_\Nav}{A_P}{\CF}$.
Similar to the proof of \\ $\AG{\Nav}{s_\Nav}{A_P}{A_{\BE}}$, BC2 will track the forward path exactly, in the reverse direction.  Since the forward
path is collision-free as guaranteed by the proof that AC2 satisfies the CF property, it immediately follows that the backtracking path is collision-free as well, i.e.,
$\AG{BC2}{s_\Nav}{A_P}{\CF}$.
}

}
	
\end{document}